\documentclass[twoside,twocolumn,9pt]{article}
\usepackage{extsizes}
\usepackage[super,sort&compress,comma]{natbib} 
\usepackage[version=3]{mhchem}
\usepackage[left=1.5cm, right=1.5cm, top=1.785cm, bottom=2.0cm]{geometry}
\usepackage{balance}
\usepackage{mathptmx}
\usepackage{sectsty}
\usepackage{graphicx} 
\usepackage{lastpage}
\usepackage[format=plain,justification=justified,singlelinecheck=false,font={stretch=1.125,small,sf},labelfont=bf,labelsep=space]{caption}
\usepackage{float}
\usepackage{fancyhdr}
\usepackage{fnpos}
\usepackage[english]{babel}
\addto{\captionsenglish}{%
  
}
\usepackage{bm}

\usepackage{mathrsfs}%
\usepackage{amsmath,amssymb,amsfonts}%
\usepackage{amsthm}%
\usepackage{textcomp}%
\usepackage{manyfoot}%
\usepackage{array}
\usepackage{droidsans}
\usepackage{charter}
\usepackage[T1]{fontenc}
\usepackage[usenames,dvipsnames]{xcolor}
\usepackage{setspace}
\usepackage[compact]{titlesec}
\usepackage{hyperref}

\usepackage{epstopdf}

\definecolor{cream}{RGB}{222,217,201}
\definecolor{green}{RGB}{0,153,51}

\begin{document}

\pagestyle{fancy}
\thispagestyle{plain}
\fancypagestyle{plain}{
\renewcommand{\headrulewidth}{0pt}
}

\makeFNbottom
\makeatletter
\renewcommand\LARGE{\@setfontsize\LARGE{15pt}{17}}
\renewcommand\Large{\@setfontsize\Large{12pt}{14}}
\renewcommand\large{\@setfontsize\large{10pt}{12}}
\renewcommand\footnotesize{\@setfontsize\footnotesize{7pt}{10}}
\makeatother

\renewcommand{\thefootnote}{\fnsymbol{footnote}}
\renewcommand\footnoterule{\vspace*{1pt}%
\color{cream}\hrule width 3.5in height 0.4pt \color{black}\vspace*{5pt}} 
\setcounter{secnumdepth}{5}

\makeatletter 
\renewcommand\@biblabel[1]{#1}            
\renewcommand\@makefntext[1]%
{\noindent\makebox[0pt][r]{\@thefnmark\,}#1}
\makeatother 
\renewcommand{\figurename}{\small{Fig.}~}
\sectionfont{\sffamily\Large}
\subsectionfont{\normalsize}
\subsubsectionfont{\bf}
\setstretch{1.125} 
\setlength{\skip\footins}{0.8cm}
\setlength{\footnotesep}{0.25cm}
\setlength{\jot}{10pt}
\titlespacing*{\section}{0pt}{4pt}{4pt}
\titlespacing*{\subsection}{0pt}{15pt}{1pt}

\fancyfoot{}
\fancyfoot[LO,RE]{\vspace{-7.1pt}\includegraphics[height=9pt]{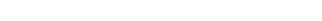}}
\fancyfoot[CO]{\vspace{-7.1pt}\hspace{13.2cm}\includegraphics{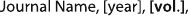}}
\fancyfoot[CE]{\vspace{-7.2pt}\hspace{-14.2cm}\includegraphics{head_foot/RF}}
\fancyfoot[RO]{\footnotesize{\sffamily{1--\pageref{LastPage} ~\textbar  \hspace{2pt}\thepage}}}
\fancyfoot[LE]{\footnotesize{\sffamily{\thepage~\textbar\hspace{3.45cm} 1--\pageref{LastPage}}}}
\fancyhead{}
\renewcommand{\headrulewidth}{0pt} 
\renewcommand{\footrulewidth}{0pt}
\setlength{\arrayrulewidth}{1pt}
\setlength{\columnsep}{6.5mm}
\setlength\bibsep{1pt}

\makeatletter 
\newlength{\figrulesep} 
\setlength{\figrulesep}{0.5\textfloatsep} 

\newcommand{\topfigrule}{\vspace*{-1pt}%
\noindent{\color{cream}\rule[-\figrulesep]{\columnwidth}{1.5pt}} }

\newcommand{\botfigrule}{\vspace*{-2pt}%
\noindent{\color{cream}\rule[\figrulesep]{\columnwidth}{1.5pt}} }

\newcommand{\dblfigrule}{\vspace*{-1pt}%
\noindent{\color{cream}\rule[-\figrulesep]{\textwidth}{1.5pt}} }

\makeatother

\twocolumn[
  \begin{@twocolumnfalse}
{\includegraphics[height=30pt]{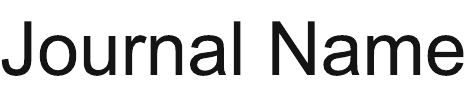}\hfill\raisebox{0pt}[0pt][0pt]{\includegraphics[height=55pt]{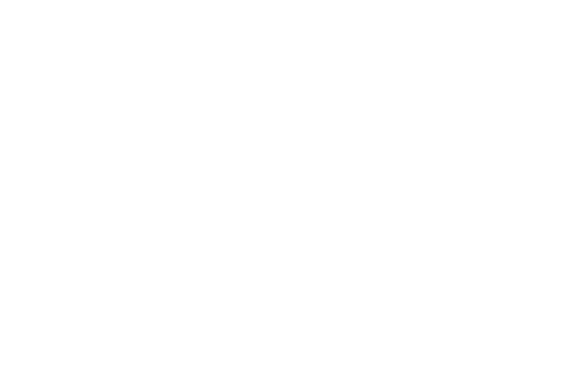}}\\[1ex]
\includegraphics[width=18.5cm]{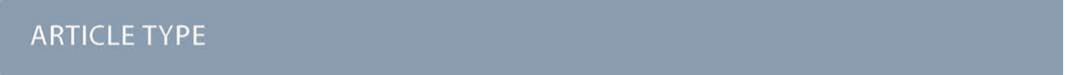}}\par
\vspace{1em}
\sffamily
\begin{tabular}{m{4.5cm} p{13.5cm} }

\includegraphics{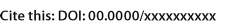} & \noindent\LARGE{\textbf{Generalization of Long-Range Machine Learning Potentials in Complex Chemical Spaces}} \\
\vspace{0.3cm} & \vspace{0.3cm} \\

 & \noindent\large{Michał Sanocki,\textit{$^{a}$}  and Julija Zavadlav $^{\ast}$\ \textit{$^{a}$}} \\ 

\includegraphics{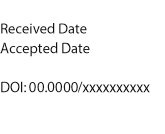} & \noindent\normalsize{The vastness of chemical space makes generalization a central challenge in the development of machine learning interatomic potentials (MLIPs). While MLIPs could enable large-scale atomistic simulations with near-quantum accuracy, their usefulness is often limited by poor transferability to out-of-distribution samples. Here, we systematically evaluate different MLIP architectures with long-range corrections across diverse chemical spaces and show that such schemes are essential, not only for improving in-distribution performance but, more importantly, for enabling significant gains in transferability to unseen regions of chemical space. To enable a more rigorous benchmarking, we introduce biased train–test splitting strategies, which explicitly test the  model performance in significantly different regions of chemical space. Together, our findings highlight the importance of long-range modeling for achieving generalizable MLIPs and provide a framework for diagnosing systematic failures across chemical space. While this study focuses on metal–organic frameworks and related systems, the proposed methodology is not limited to this class of materials and may inform the design of more robust and transferable MLIPs in other systems."
}\\

\end{tabular}

 \end{@twocolumnfalse} \vspace{0.6cm}
  ]

\renewcommand*\rmdefault{bch}\normalfont\upshape
\rmfamily
\section*{}
\vspace{-1cm}


\footnotetext{\textit{$^{a}$~Multiscale Modeling of Fluid Materials, Department of Engineering Physics and
Computation, TUM School of Engineering and Design, Technical University of Munich, Germany; E-mail: julija.zavadlav@tum.de}}





\section*{Introduction}

One of the largest issues in data-driven chemical modeling is the generalizability of the developed methods over the vast chemical space \cite{Pyzer-Knapp2019,Gallegos2021-kv, Segal2025-hb}. In fact, even the chemical space of only small organic molecules has been estimated to encapsulate around $10^{60}$ possibilities \cite{https://doi.org/10.1002/(SICI)1098-1128(199601)16:1<3::AID-MED1>3.0.CO;2-6}. This challenge is not limited to computational methods, as experimental methods also struggle with the sheer enormity of the possible molecular space \cite{LAVECCHIA2024104133}. Paradoxically, the difficulty in exploring chemical space has often been the reason for adopting data-driven approaches  \cite{Betinol2023-vk, Schwaller2019}. Navigating such a vast chemical space poses a fundamental challenge for predictive modeling: no matter how large a dataset is, it will inevitably cover only a minute fraction of possible chemistries. As a result, the core problem for universal models is not whether they can interpolate within known regions of chemical space, but whether they can generalize to unseen regions. 

This is especially problematic for the development of machine learning interatomic potentials (MLIPs), which have become increasingly popular in recent years due to their ability to deliver near-DFT accuracy at a fraction of the computational cost \cite{Anstine2023},  enabling simulations of larger systems and longer timescales that would otherwise be unfeasible \cite{Eckhoff2019-jl, unkeMachineLearningForce2021a}.
The increasing popularity of MLIPs has also led to a surge in new architectures, with graph neural networks (GNN) emerging as a particularly promising approach \cite{Batzner2022}. Further advances, such as equivariant GNNs (ensuring the preservation of physical symmetries) and message-passing, have significantly improved the viability of these models \cite{Musaelian2023}. This has made MLIPs a practical alternative to both classical force fields and quantum methods. However, their usefulness is ultimately constrained by the issue of generalization to out-of-distribution samples \cite{Cui2025-yg}. This limitation arises from not only the vastness of chemical space but also the conformational diversity of individual molecules. At the same time, there has been a growing interest in developing universal or foundational MLIPs\cite{batatia2024foundationmodelatomisticmaterials,wood2025umafamilyuniversalmodels,Kovacs2025-za, benedini2025universalmachinelearningpotential}. This trend is motivated by the desire to cover broader regions of chemical space with a single model, making generalizability even more important.


\begin{figure*}[ht]
    \centering
    \includegraphics[width=0.65\linewidth]{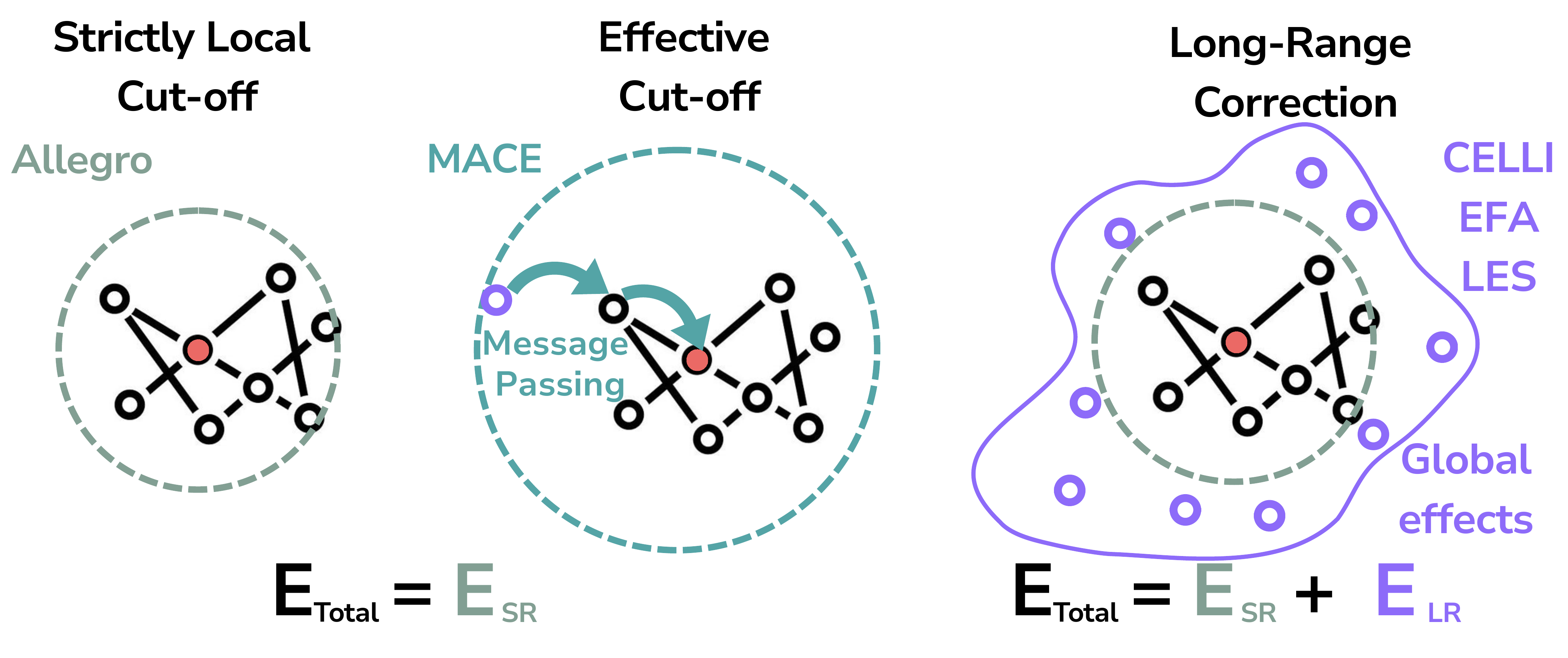}
    \caption{Difference between strictly local, message passing, and long-range methods in MLIPs. We refer to the cutoff extended by message passing layers as the effective cutoff to differentiate it from the cutoff present in strictly local models.
}
    \label{fig:LR_influence}
\end{figure*}
Achieving a truly generalizable MLIP would require capturing different types of interactions without overestimating any single one. Typically the total energy of a system is decomposed into $E_{\text{SR}}$ (short-range) and $E_{\text{LR}}$ (long-range) contributions:
\begin{equation}
    E_{\text{total}} = E_{\text{SR}} + E_{\text{LR}}.
\end{equation}
Due to computational constraints, MLIPs can only access interactions within a finite effective cut-off radius. Thus, they are likely to overestimate short-range interactions to compensate for the missing or incorrectly modeled long-range contributions, especially in strictly local models such as Allegro \cite{Musaelian2023}, which only exchange information within a short cut-off \cite{Batzner2022} (see Figure \ref{fig:LR_influence}). This would suggest that such models may overfit to the training data, thereby reducing their generalizability. 

The issues with strictly local cutoffs can also be partially offset by message-passing neural networks (MPNNs), which gained significant popularity partially due to their longer effective receptive fields. However, they face challenges related to parallelization and scalability \cite{Anstine2023}, particularly in the context of molecular dynamics simulations of large systems, due to the growth of their receptive fields \cite{Musaelian2023}, bottlenecks in per-layer communication \cite{xia2025efficientparallelizationmessagepassing}, and layer-wise synchronization barriers \cite{besta2023paralleldistributedgraphneural}. These limitations have motivated the development of strictly local models, which are well suited for incorporating long-range correction schemes due to their lower computational cost \cite{Batzner2022}. Additionally, increasing the number of propagation layers beyond a certain point leads to diminishing returns and eventually feature collapse \cite{Keriven2022}. We hypothesize that separate modeling of short- and long-range energy contributions may improve performance by encouraging more balanced representations of both interaction types. While such separation does not by itself guarantee improved generalization, it may mitigate overfitting arising from compensatory adjustments within the short-range component.

Recently, many different approaches to long-range modeling have been proposed with solutions ranging from charge equilibration schemes \cite{CENT, Ko2021,fuchs2025learningnonlocalmolecularinteractions} to charge-independent methods such as self-consistent neural networks \cite{SCFNN}, reciprocal space transformations \cite{kosmala_ewaldbased_2023}, Gaussian multipliers \cite{ssp9-7s81}, attention-based architectures \cite{frank2024euclideanfastattentionmachine}, and methods based on Ewald summation \cite{Cheng2025-dw}. It is worth noting that, despite their proven viability, many foundational models do not incorporate long-range schemes \cite{wood2025umafamilyuniversalmodels,Kovacs2025-dd}, which may explain why they often struggle to predict experimental measurements \cite{mannan2025evaluatinguniversalmachinelearning}.

The challenges with chemical diversity and long-range effect modeling outlined above are particularly pronounced in the context of metal–organic frameworks (MOFs). As their modular architecture allows for precise control over porosity, surface area, and chemical functionality, making them highly tunable for a wide range of applications, including gas storage, catalysis, separations, and sensing \cite{EXP, Burner2020-xp}. Unlike traditional porous materials such as zeolites, MOFs offer exceptional structural diversity; tens of thousands of variants have already been synthesized \cite{EXP2}, and computational design opens the door to virtually limitless hypothetical structures \cite{QMOF,Kang2023-jj}. This vast chemical and configurational space makes it essentially impossible to identify optimal materials for specific applications through empirical methods alone \cite{ARCMOF, D3SC05612K, Sriram2024-sf}. Computational methods are therefore essential to accelerate MOF discovery and to probe their behavior under working conditions, using methods such as classical MD \cite{Islamov2023-ui}, DFT \cite{DAVIS2025113537}, and grand canonical Monte Carlo \cite{TAO2022e00383,https://doi.org/10.1002/aoc.7199}.

However, classical modeling techniques face several limitations in their applicability to MOF modeling, as they face trade-offs between accuracy and computational efficiency, making accurate large-scale MOF simulations unfeasible \cite{Vandenhaute2023-qc, Sriram2024-sf}. Therefore, MOFs present an ideal application for MLIPs, as classical force fields often lack the accuracy required to capture the complex interactions and are very difficult to parameterize, while DFT is too computationally expensive to model structures with large unit cells and at longer time scales \cite{Eckhoff2019-jl, unkeMachineLearningForce2021a, Zhang2023-kc}, thus making MLIPs a perfect candidate for large scale MOF simulations. As a result, a wide range of studies have employed MLIPs to investigate the chemical and mechanical properties of MOFs \cite{Vandenhaute2023-qc, Eckhoff2019-jl, D3NR05966A, Sharma2024-gn, Luo2024-yl,Kancharlapalli2021-wj}. However, two issues limit the usefulness of MLIPs. First, they often struggle to generalize to out‑of‑distribution samples, which confines their applicability to a narrow region of chemical space \cite{Cui2025-yg}. 
Second, long‑range effects and electrostatics, which are particularly challenging to model for MLIPs \cite{unkeMachineLearningForce2021a}, can significantly affect MOF behavior \cite{Thaler2024, 10.1063/5.0054874}. Additionally, foundational models also struggle with modeling MOFs, often failing to outperform simple classical force fields \cite{kraß2025mofsimbenchevaluatinguniversalmachine}, suggesting that large training sets might not be enough to achieve truly universal MLIPs. All of these factors make MOFs an ideal target for our study, as they are highly relevant to experimental and computational chemists, have been previously investigated using MLIPs, and exhibit a vast and complex chemical space, providing an ideal test case for evaluating model generalizability  across chemical space.

In this work, we test the generalizability of three widely used baseline architectures: DimeNet++ \cite{gasteigerFastUncertaintyAwareDirectional2022}, MACE \cite{NEURIPS2022MACE}, and Allegro \cite{Musaelian2023}, on diverse chemical spaces defined by subsets of QMOF \cite{QMOF}, ODAC25 \cite{sriram2025opendac2025dataset}, and OMOL25 \cite{levine2025openmolecules2025omol25} datasets. 
We perform a direct comparison of different long-range correction schemes; thus, we test two recently introduced frameworks in combination with different baseline models: the Charge Equilibration Layer for Long-range Interactions (CELLI), recently introduced by the authors \cite{fuchs2025learningnonlocalmolecularinteractions}, and Euclidean Fast Attention (EFA) \cite{frank2024euclideanfastattentionmachine}.
 We show that such corrections not only allow cheaper models to achieve state-of-the-art accuracy but are also essential for improving generalizability across chemical space, even in MPNNs. Unlike most prior works \cite{Kovacs2021-uq,Liu2023-ca}, which have mainly focused on conformational generalization, we center our analysis on chemical diversity. Furthermore, we demonstrate that partial charges cannot be inferred without training on reference partial charge labels in the case of challenging datasets such as the ones tested in this work. This is also the case for the Latent Ewald Summation (LES) \cite{Cheng2025-dw} framework, which recently claimed the opposite \cite{Cheng2025-dw,kim2025universalaugmentationframeworklongrange, King2025-gt, zhong2025machine}. In these challenging environments, models trained with CELLI based on reference charges consistently produce physically meaningful results, highlighting that leveraging accurate charge information remains critical for developing truly generalizable long-range MLIPs.


\section*{Methods}

\subsection*{Experiment Design}

To evaluate generalizability, we use three datasets: QMOF \cite{QMOF}, ODAC25 \cite{sriram2025opendac2025dataset}, and a metal-complex subsplit of OMOL25 \cite{levine2025openmolecules2025omol25}. The QMOF dataset contains MOFs with up to 500 atoms per unit cell in their ground state, which is particularly suitable here since our focus is on generalizability across chemical space. By contrast, ODAC25 and OMOL25 include non-ground-state structures. To ensure methodological consistency, maintain acceptable computational cost for training a large number of MLIPs, and focus on chemical space generalization, we constructed subsplits containing only the lowest-energy molecules. Although OMOL25 does not include MOF structures but only metal-organic complexes, we incorporated it due to the lack of alternative MOF datasets. To our knowledge, OMOL25 also lacks ground-state structures, so we selected the lowest-energy conformer for each molecule (see Supporting Information  section 2 for details on subsplit construction). Likewise, for ODAC25, we used up to the 10 lowest-energy conformations per molecule, otherwise, the resulting subsets would be too small to train or evaluate models reliably. The resulting subsets comprise 76,525 unique molecules from OMOL25 (up to 350 atoms), 20869 MOFs from ODAC25 (up to 616 atoms), and the full QMOF dataset with 20,407 MOFs (up to 500 atoms).

To evaluate how well different MLIPs generalize to out-of-distribution samples, we consider four evaluation strategies. First, we use a previously introduced method, where the model is trained on a subset of structures with 100 or fewer atoms and then tested on a subset of larger molecules \cite{Thaler2024}. Since larger molecules are likely to be significantly different from those in the smaller subset. Secondly, we introduce two additional biased train-test split methods: cluster and maximal separation, to systematically investigate differences in model generalization to unseen regions of chemical space. Lastly, a regular random split was used as a comparison to biased split methods.

To create our biased split methods, we employ SOAP (Smooth Overlap of Atomic Positions) descriptors \cite{PhysRevB.87.184115}, which combine radial and angular information into rotationally invariant atomic features. First, we calculate SOAP descriptors of each atom and average them to create a global similarity measure \cite{DScribe}. Unfortunately, this vector is uninterpretable; however, it allows us to define an architecture-independent descriptor space for molecules, enabling the investigation of MLIP's performance in out-of-distribution samples. The SOAP average similarity kernel has also been previously utilized by \citeauthor{QMOF} on the QMOF dataset to identify local trends in feature space and for band gap predictions, achieving good results \cite{QMOF}, and demonstrating SOAP's ability to produce a meaningful representation of the MOF's atomic environment.

The \textbf{cluster} method groups molecules into structurally similar clusters using K-Means clustering on their SOAP descriptors. A subset of clusters is then randomly selected for the training set, while the remainder forms the test set. This enforces a structural distinction between splits.

The \textbf{maximal separation} method computes pairwise cosine similarities between descriptor vectors \textbf{A} and \textbf{B}
\begin{equation}
    \text{Similarity}(\mathbf{A}, \mathbf{B}) = \frac{\mathbf{A} \cdot \mathbf{B}}{\|\mathbf{A}\| \|\mathbf{B}\|}.
\end{equation}
Starting with a random training data sample, the maximal separation method iteratively adds a candidate to the test dataset. The selected candidate $\hat{\textbf{A}}$ has minimal similarity to training data samples \textbf{B}
\begin{equation}
    \hat{\mathbf{A}}  = \arg\min_{\mathbf{A} \in \text{Remaining}} \left( \max_{\mathbf{B} \in \text{Train}} \text{Similarity}(\mathbf{A}, \mathbf{B}) \right).
\end{equation}

 The proposed biased split methods evaluate distinct aspects of generalizability: the small–large split assesses performance on cells of varying sizes; the maximal separation method measures performance on a subset maximally different from the training set to stress-test MLIP; and the cluster method examines generalization to distinct structural families by partitioning chemical space into structurally coherent clusters. This addresses a critical limitation of MLIPs: available datasets likely under-sample the vast chemical space, creating biases that may lead to overoptimistic performance evaluation and, as a result, limit their applicability to only its narrow part.

\begin{figure*}[ht]
    \centering
    \includegraphics[width=0.7\linewidth]{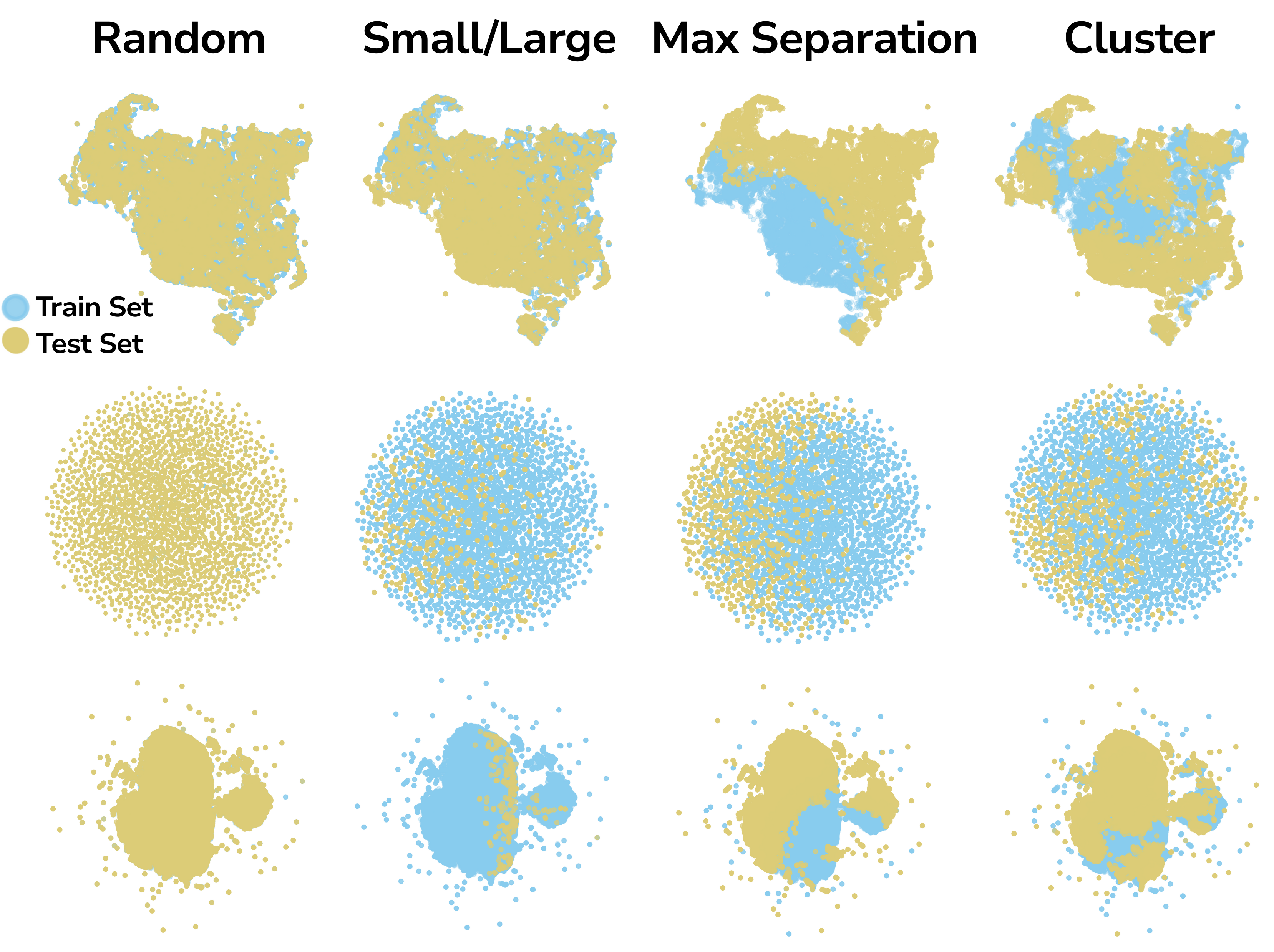}
    \caption{Overview of the max separation, cluster, size-based splits, and random for QMOF (top), ODAC25 (middle), and OMOL25 (bottom). Each split is visualized on the UMAP dimensionally reduced chemical space. In both cases, the goal is to create a train–test split that introduces significant differences between the subsets to account for the inherent bias in the available chemical space and to establish a stress test–like scenario for evaluating the generalizability of MLIPs. Here we show one split for each dataset, while the remaining splits are reported in the Supplementary Information section 2.
}
    \label{fig:over_split}
\end{figure*}

To ensure a sufficiently large test set for meaningful generalizability analysis, we allocated 50\% of the OMOL25 and QMOF datasets, and 30\% of the ODAC25 dataset, to testing. In all cases, the validation set was drawn from the training data, comprising 10\% of its datapoints. To account for random effects introduced by the splitting process, we generated each split three times with different seeds for the ODAC25 and QMOF datasets, and only once for OMOL25 due to higher computational cost. Then, to visualize how different split methods divide given chemical space, we applied Uniform Manifold Approximation and Projection (UMAP) \cite{sainburg2021parametric}, a nonlinear dimensionality reduction technique that preserves both local and global structure, and has been often used to visualize chemical space~\cite{Orlov2024-rn,SOSNIN2025104392, Boldini2024-ju}. Figure \ref{fig:over_split} shows that the proposed biased splitting methods, and especially the maximal separation method, produce splits that are significantly different (at least in the SOAP descriptor space) and can serve as a challenging benchmark for evaluating the generalizability of MLIPs.

The SOAP descriptors  were generated using the Dscribe package\cite{DScribe} and parameters provided by \citeauthor{QMOF}\cite{QMOF}. The maximal separation method is computationally expensive and scales poorly with dataset size, as it requires computing similarity  between all members of the training set and remaining molecules. For this reason, in the OMOL25 dataset, a simplified version was used, where similarity was computed only between subsets of 1000 molecules in the training set and the remaining molecules. For the cluster method, the number of clusters was set to 20, and 50\% (70\% in ODAC25) were used for training.

\subsection*{Architectures and Long-Range Corrections}

As baseline models, we consider DimeNet++ \cite{gasteigerFastUncertaintyAwareDirectional2022}, MACE \cite{NEURIPS2022MACE}, and Allegro \cite{Musaelian2023}, which represent distinct design paradigms: Allegro is an equivariant, strictly local, edge-centric model; MACE is an equivariant, message-passing, node-centric model; and DimeNet++ is an invariant message-passing architecture that incorporates explicit three-body interactions. In order to show that the improvements resulting from the introduction of long-range methods cannot be achieved by simply increasing the effective cutoff via additional message passing steps, we trained an additional baseline model with two additional (four in total) message passing layers, which we refer to as the MACE-MP4 model. MACE and Allegro models also included CELLI \cite{fuchs2025learningnonlocalmolecularinteractions} and EFA \cite{frank2024euclideanfastattentionmachine} long-range corrections schemes. Although both EFA and CELLI have shown effectiveness at modeling long-range interactions \cite{frank2024euclideanfastattentionmachine, fuchs2025learningnonlocalmolecularinteractions}, they are based on fundamentally different design principles. Specifically, EFA relies on attention mechanisms to learn global representations of chemical systems, whereas CELLI is grounded in physics-based design and dynamically redistributes charge to account for long-range interactions and charge transfer.

These differences make them ideal candidates for this study, as they allow for a comparison of the advantages and disadvantages of physics-inspired versus purely AI-driven approaches. In principle, physics-based models are expected to generalize better due to their grounding in physical theory \cite{thurlemannRegularizedPhysicsGraph2023}. However, imposing such constraints may also limit the expressiveness of the model \cite{Shaidu2024}. On the other hand, data-driven methods often offer greater flexibility but are prone to overfitting and may generalize poorly to out-of-distribution samples\cite{Cui2025-yg}. In addition, when fitted to partial charges, CELLI can only model electrostatic interactions and charge transfer, whereas EFA can represent all long-range effects, including van der Waals interactions. To ensure comparability, the same hyperparameters were used for each architecture across different splits.

\subsection*{CELLI}
The Charge Equilibration Layer for Long-range Interactions (CELLI) is a model-agnostic component that enables the integration of non-local electrostatics into local equivariant GNN architectures. CELLI follows the charge equilibration (Qeq) framework, computing partial charges based on environment-dependent electronegativities, species-dependent hardnesses, and charge radius \cite{fuchs2025learningnonlocalmolecularinteractions}. Electronegativity values are predicted per node by aggregating local scalar edge features using an MLP-based mechanism. Hardness and atomic radius values are derived from species embeddings, with radii initialized from covalent radii and scaled by a learned, positive factor. 

Those parameters are then used to redistribute charge based on the following energy minimization problem. The total Qeq energy is defined as
\begin{align}
 \alpha_{ij} = \frac{1}{\sqrt{2}}(\gamma_i^2 + \gamma_j^2)^{-1/2}\\
    U_\text{Coul}(\bm R, \bm Q) = \sum_{i}^N\sum_{j>i}^N\frac{\operatorname{erf}(\alpha_{ij} r_{ij})}{r_{ij}}Q_iQ_j + \sum_{i=1}^N\frac{2\alpha_{ii}}{\sqrt{\pi}}Q_i^2\\
U_\text{Qeq}(\bm R, \bm Q) = U_{\text{Coul}}(\bm R, \bm Q) + \sum_{i=1}^N\left[ \chi_i Q_i + \frac{J_{ii}}{2} Q_i^{2} \right]
\end{align}
where the final term captures the atomic contributions through previously computed electronegativities $\chi_i$, atomic radii $\gamma_i$, and chemical hardnesses $J_{ii}$ \cite{qeq}. The minimum of $U_\text{Qeq}$ is the solution of the linear system
\begin{align}
\left[\left.\frac{\partial^2 U_\text{Coul}}{\partial Q_i\partial Q_j}\right|{\bm R} + J{ii} \right] Q_j = -\chi_i
\end{align}
under the charge conservation constraint.

Once charges are obtained, they are embedded into the GNN by generating charge-dependent feature vectors via a second MLP, which are combined with existing latent scalar features through a residual update. This allows downstream layers to incorporate non-local information while maintaining the local structure of the network. Although CELLI is model agnostic and can be utilized with multiple frameworks, it's best suited for strictly local models such as Allegro \cite{Musaelian2023}, as the benefits of long-range scheme correction are larger compared to message passing architectures. However, here we also consider an MACE implementation. Regardless of the MLIP architecture used, the long-range electrostatic interactions could be partially accounted for by the long-range contribution and partially by the short-range contribution to the total energy. However, when CELLI is trained on partial charges, electrostatic interactions are assumed to be entirely accounted for by the long-range contribution. In other words, the electrostatic interactions are not double-counted.

\subsection*{EFA}

Euclidean Fast Attention (EFA) encodes atomic spatial information into feature representations by combining distance-aware modulation with rotational invariance through integration over the unit sphere. EFA builds on Euclidean Rotary Positional Encoding (ERoPE)\cite{SU2024127063}, which maps 3D positions into complex-valued feature vectors using frequency-based phase shifts \cite{frank2024euclideanfastattentionmachine}. 

Specifically, given a position vector $\mathbf{r} \in \mathbb{R}^3$, a feature vector $\mathbf{x}$, a frequency $\omega \in \mathbb{R}$, and a unit vector $\mathbf{u} \in S^2$, ERoPE modulates the feature as
\begin{equation}
\phi_{\mathbf{u}}(\mathbf{x}, \mathbf{r}) = \mathbf{x} \cdot e^{i \omega \mathbf{u} \cdot \mathbf{r}},
\end{equation}
where the dot product $\mathbf{u} \cdot \mathbf{r}$ projects the position onto the direction $\mathbf{u}$. To ensure rotational invariance, EFA averages over all directions $\mathbf{u}$ on the unit sphere $S^2$, resulting in the attention mechanism
\begin{equation}
\text{EFA}(\mathbf{X}, \mathbf{R})_m = \frac{1}{4\pi} \int_{S^2} \phi_{\mathbf{u}}(\mathbf{q}_m, \mathbf{r}_m)^\top \sum_{n=1}^N \phi_{\mathbf{u}}(\mathbf{k}_n, \mathbf{r}_n) \mathbf{v}_n^\top \, d\mathbf{u},
\end{equation}
where $\mathbf{q}_m$, $\mathbf{k}_n$, and $\mathbf{v}_n$ are the query, key, and value vectors for atoms $m$ and $n$, respectively, and $\mathbf{r}_m$, $\mathbf{r}_n$ are their 3D coordinates. 

By integrating over all directions on the unit sphere, the resulting attention weights depend only on interatomic distances, ensuring rotational invariance. To support directional reasoning, EFA incorporates a tensor product with spherical harmonics, allowing the mechanism to operate on equivariant inputs. This enables the network to model both long-range interactions and geometric relationships without requiring explicit neighbor cutoffs, making it well-suited for large atomistic systems.

Originally, the EFA scheme was proposed for a generic MPNN architecture, which consisted only of node features and a message passing step, which strongly limits its applicability. Thus, there is a need to adapt this scheme to popular frameworks. Unlike in the network utilized by \citeauthor{frank2024euclideanfastattentionmachine}, there are several ways in which EFA can be integrated into MACE or Allegro. However, the detailed analysis of the optimal EFA implementation is outside of the scope of this study, and in this section, we provide a basic description of our integration. 

The EFA block takes as input the current node embeddings and atomic positions. To integrate it with MACE, the output is then fused with the node features before being passed into subsequent MACE interaction layers. By inserting EFA at configurable points in the message passing stack, the architecture captures long-range geometric dependencies prior to or between equivariant tensor-product updates. This design enables the network to blend fast, global spatial awareness with deep local equivariant reasoning, allowing for flexible integration of EFA alongside or in place of conventional MACE message passing without disrupting equivariance or scalability.

In the Allegro integration, EFA features are combined with the original species embeddings and linearly projected into a refined feature space just before the Allegro tensor product layer (initial invariant scalar latent features in the first MLP only utilize original species embeddings). This insertion point ensures that EFA-enhanced node-level information flows into the pairwise feature prediction pipeline of Allegro. EFA was initially implemented for MPNNs, however, message passing is not fully required and can be combined with a strictly-local Allegro architecture. 



\subsection*{LES}

Unlike charge-equilibration approaches, LES does not solve a redistribution problem. Instead, LES learns latent electrostatic charges directly from local atomic descriptors and uses them to compute a long-range energy contribution through an Ewald formulation \cite{Cheng2025-dw}. These charges are not constrained to match reference partial charges and are optimized solely through the global energy and force loss terms. As a result, they do not conserve the total charge. For periodic systems, the reciprocal-space Ewald term is used:
\begin{align}
U_{\mathrm{lr}}(\mathbf{R}, \mathbf{q})
&=
\frac{1}{2 \varepsilon_0 V}
\sum_{0 < |\mathbf{k}| < k_c}
\frac{e^{-\sigma^2 |\mathbf{k}|^2 / 2}}{|\mathbf{k}|^2}
\left| S(\mathbf{k}) \right|^2, \\
S(\mathbf{k})
&=
\sum_{i=1}^N
q_i \, e^{i \mathbf{k} \cdot \mathbf{r}_i},
\end{align}
where $V$ is the simulation cell volume and $\sigma$ is a Gaussian smearing parameter.

For isolated systems, LES uses a screened direct Coulomb sum:
\begin{align}
U_{\mathrm{lr}}(\mathbf{R}, \mathbf{q})
&=
\frac{1}{8\pi\varepsilon_0}
\sum_{i=1}^N
\sum_{j>i}^N
\frac{q_i q_j}{r_{ij}}
\left[
1 - \operatorname{erf}\!\left(
\frac{r_{ij}}{\sqrt{2}\sigma}
\right)
\right].
\end{align}

LES also enables computation of Born effective charge tensors, which can be used for simulations of systems under an electric field \cite{King2025-gt}(for details, see Supplementary Information section 9). After computing the long-range energy, LES includes electrostatic information into the model only through its total energy contribution. No additional solvers, constraints, or charge-conservation steps are utilized. In this work, we utilized the MACE \cite{NEURIPS2022MACE} \cite{cheng2024cartesian} integrations of LES due to the high computational cost of CACE (for details on reproducibility of previous results, see Supplementary Information section 8).

\subsection*{Total Charge Embeddings}

First, we extend the standard MACE node-embedding layer to incorporate a global conditioning on total charge alongside atomic species. In addition to the species embedding, a separate embedding is learned and projected into the same scalar feature space and broadcast across all nodes. The species and charge embeddings are fused and refined through an MLP. This modification injects global charge information into every node before message passing, enhancing expressivity without affecting the symmetry guarantees of the core MACE architecture.

For Allegro, we combine learned embedding of total charge with the species embeddings and linearly project into a refined feature space just before the Allegro tensor product layer (similarly to the EFA integration initial invariant scalar latent features in the first MLP, only utilize original species embeddings). Although Allegro is strictly local, the global charge embedding provides a simple mechanism to modulate all node features consistently based on system-level charge, enhancing expressivity without modifying the underlying equivariant structure.

\subsection*{Model Training}

All baseline, CELLI and EFA models were trained in JAX using chemtrain \cite{fuchs2024chemtrainlearningdeeppotential}, adapted JAX-MD \cite{schoenholzJAXFrameworkDifferentiable2021} packages together with JAX-compatible implementations of Allegro \cite{marioAllegrojax}, DimeNet++ \cite{thalerLearningNeuralNetwork2021, gasteigerFastUncertaintyAwareDirectional2022}, and MACE \cite{NEURIPS2022MACE,marioMACEjax} via the Force Matching method \cite{ercolessiInteratomicPotentialsFirstPrinciples1994, fuchs2024chemtrainlearningdeeppotential}.
For the OMOL25 and QMOF datasets, models with CELLI were first pretrained using only charges, and subsequently trained on charges, forces, and energies. Since the ODAC25 dataset does not include force information, it was excluded from this training procedure. For details on training, data preprocessing, and hyperparameters, see Supplementary Information (sections 3-5). We restrict our evaluation to molecules containing only species that appear at least 10 times in the training set, as it is unrealistic to expect accurate modeling of rare elements. Species that occurred only in the test set were also discarded. 
Models with LES were trained using MACE \cite{NEURIPS2022MACE, ilyes_batatia_2025_16748079} and LES \cite{lesfit} PyTorch implementation.

\section*{Results}

\subsection*{Increased Generalization of Long-Range Models}

\begin{figure*}[ht]
    \centering
    \includegraphics[width=0.7\linewidth]{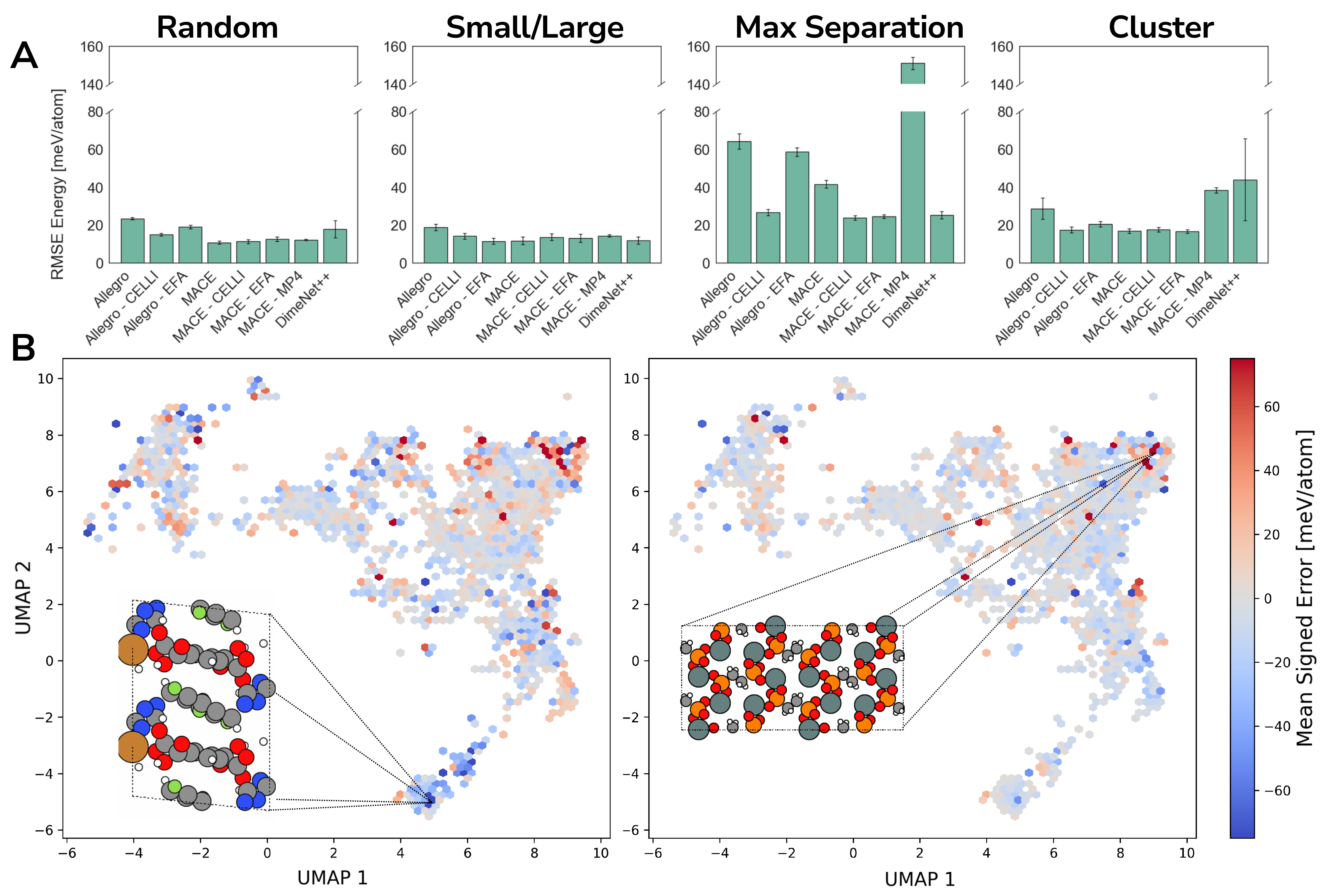}
    \caption{ \textbf{Long-Range schemes increase generalizability of MLIPs.}
    \textbf{A} Average model performance across maximum separation, cluster, size, and random data splits on the QMOF dataset measured by Root Mean Squared Error (RMSE). Each model was trained three times with different initializations. The QMOF dataset provides only energies and no force labels.
    \textbf{B}
    Average error on the QMOF test space of the max separation split for (short-range) Allegro (left) and (long-range) Allegro-CELLI (right). The highlighted molecules represent areas of chemical space with similar biases. Detailed results and plots for all models are available in the Supporting Information section 7.}
    \label{fig:space_error}
\end{figure*}

Our results show that the integration of long-range correction schemes has a significant effect on both MACE and Allegro and is necessary for the latter to achieve SOTA performance on the QMOF dataset (see Figure \ref{fig:space_error}). For Allegro, this improvement in Root Mean Squared Error (RMSE) is noticeable even in the simplest benchmark (random split), which aligns with our expectations, as Allegro is strictly local and therefore struggles to model systems where long-range interactions can be significant. However, incorporating EFA into Allegro yields a smaller performance gain compared to CELLI, especially for the cluster and maximal separation splits, suggesting that EFA does not generalize as well to out-of-distribution samples as CELLI. This difference may be influenced by two factors: (1) the inherent generalizability of physics-based models, and (2) the large chemical diversity of the QMOF dataset relative to its size (only 20,387 samples), as EFA may outperform CELLI in larger datasets.

For MACE trained on the random split, the effect of adding long-range schemes is minimal and comparable to statistical noise; however, in more challenging test cases (maximum separation split), the improvement becomes evident even for MACE, as it fails to generalize to out-of-distribution test sets. 
Although CELLI and EFA models also struggle with biased splits, they perform significantly better than baseline models, suggesting that long‐range corrections are crucial for achieving robust generalization. Our results also show that increasing the number of message passing steps can lead to overfitting and a decrease in performance compared to the baseline model. For example, in the cluster and maximum separation splits, an increase in the number of message passing steps more than doubles the RMSE, showing that the addition of separate long-range corrections is advantageous. 

Interestingly, DimeNet++, which obtains reasonable results in the random, size, and maximum separation splits, fails to generalize under cluster split. This shows that the proposed benchmarks probe different aspects of generalizability and can be used in future studies focusing on the development of long-range correction schemes. Our results also support our earlier hypothesis on the necessity of inclusion of a dedicated mechanism for modeling long-range interactions. The issues with generalizability are particularly significant for MACE, as foundational and universal models have been developed using this architecture \cite{Kovacs2025-za, batatia2024foundationmodelatomisticmaterials}. However, our results show that MACE without any long-range correction does not generalize well to out-of-distribution samples, suggesting it may not be suitable for this task.
\subsection*{Modeling Charged Systems}
\begin{figure*}[ht]
    \centering
    \includegraphics[width=0.7\linewidth]{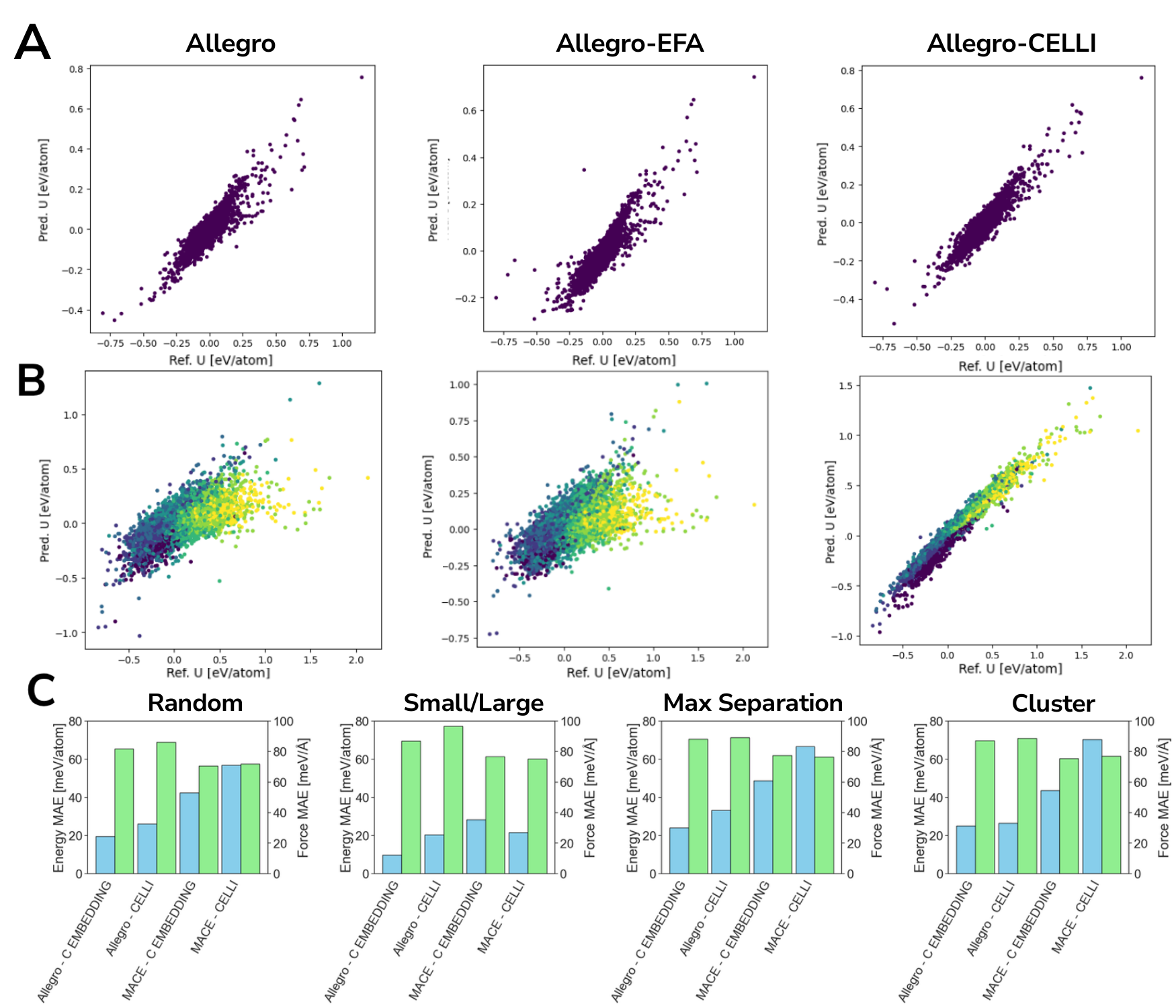}
    \caption{ \textbf{Accuracy on Charged vs. Neutral Systems}
Parity plots for three Allegro models with and without long-range corrections trained on only neutral (\textbf{A}) and both neutral and charged molecules (\textbf{B}) from the OMOL25 dataset. Colors correspond to the total charge. Baseline Allegro and Allegro-EFA models exhibit a significant increase in error when trained on charged samples.  
    \textbf{C} Model performance across maximal separation, cluster, size, and random data splits on the OMOL25 dataset measured by Mean Absolute Error (MAE). Blue bars correspond to energy MAE, while green bars correspond to force MAE. “C - EMBEDDING” corresponds to models with an additional embedding for total charge; see Methods for details.}
    \label{fig:omol_fig}
\end{figure*}

The biased dataset splits introduced in this study also allow us to visually investigate the generalizability of MLIPs across chemical space. To illustrate this, we visualized the prediction error of each model on the test portion of the max separation split of the QMOF dataset (see Figure~\ref{fig:space_error} \textbf{B}). Interestingly, although the Allegro–CELLI model achieves significantly better overall accuracy, both the baseline Allegro and Allegro–CELLI exhibit similar regions of bias across chemical space. The magnitude of the error is substantially larger for the baseline Allegro model, and only a few regions show qualitatively different types of error (i.e., overestimation versus underestimation). This observation suggests that certain areas of chemical space are inherently challenging for both models. Possible explanations include intrinsic difficulty in modeling complex structures, inaccuracies within the QMOF dataset, or undersampling of specific chemical motifs. Nevertheless, the model incorporating long-range corrections performs uniformly better, highlighting that such corrections are essential for improving the generalizability of MLIPs. It should also be noted that near-zero average error does not necessarily mean that the error is  smaller, only that it does not exhibit bias.


In our next benchmark, we utilized the OMOL25 dataset to evaluate the performance of different schemes on charged molecules. First, we trained MACE and Allegro on the neutral molecules of our OMOL25 subset, with and without long-range schemes. While performance on the neutral subset is not strongly affected by CELLI, the baseline versions of both MACE and Allegro are degenerate with respect to total charge, meaning they cannot distinguish between molecules with different charge states and assign identical energies to species that differ only in charge. This limitation becomes critical once multiple charge states are present (see Figure~\ref{fig:omol_fig}); therefore, we limited our subsequent analysis to the charge-dependent models.

To enable a fair comparison for CELLI, we added total charge embeddings to baseline Allegro and MACE (for details, see Methods). Interestingly, models with total charge embeddings perform slightly better compared to CELLI in the majority of cases. This demonstrates that CELLI can effectively act as a total charge embedding scheme, while also offering two key practical advantages over explicit embeddings: 1) it provides long-range capabilities, which are likely to be relevant in simulations of larger systems, and 2) although accounting for nonzero total charges is necessary for training, MD simulations are typically conducted under neutral conditions, meaning those embeddings would not be used outside of training. Additionally, total charge embeddings can also be combined with all long-range methods, including CELLI; hence, they should not be viewed as a replacement, but rather as a potential extension. The integration of these embeddings into long-range schemes is, however, beyond the scope of this study.

\subsection*{Inferring Charges from Forces and Energy}

\begin{figure*}[ht]
    \centering
    \includegraphics[width=0.9\linewidth]{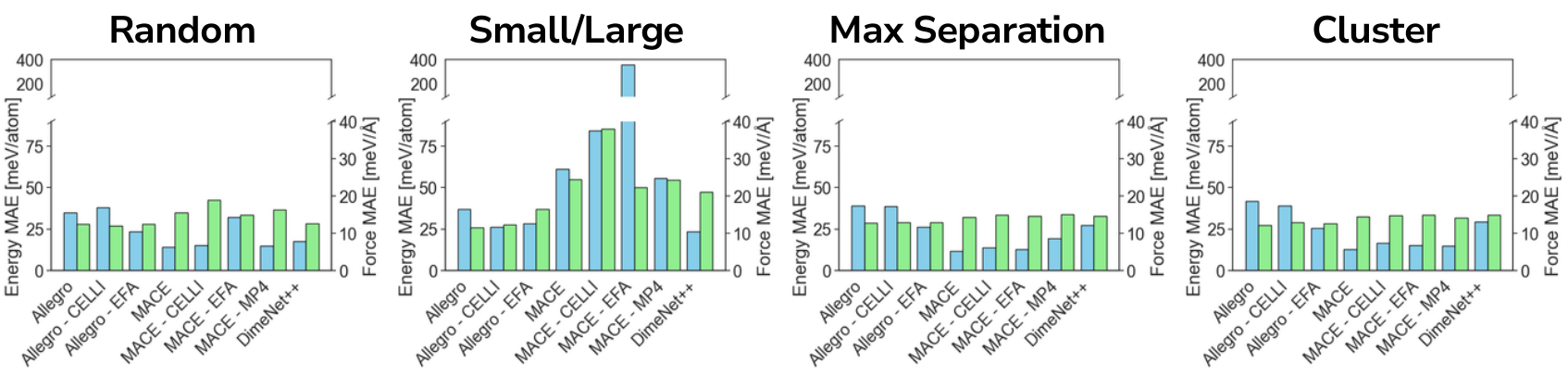}
    \caption{ \textbf{Long-range modeling without reference charges.}
    Model performance across maximal separation, cluster, size, and random data splits on the ODAC25 dataset.  Blue bars correspond to energy MAE, while green bars correspond to force MAE.  }
    \label{fig:ODAC}
\end{figure*}
\begin{figure*}[h]
    \centering
    \includegraphics[width=0.7\linewidth]{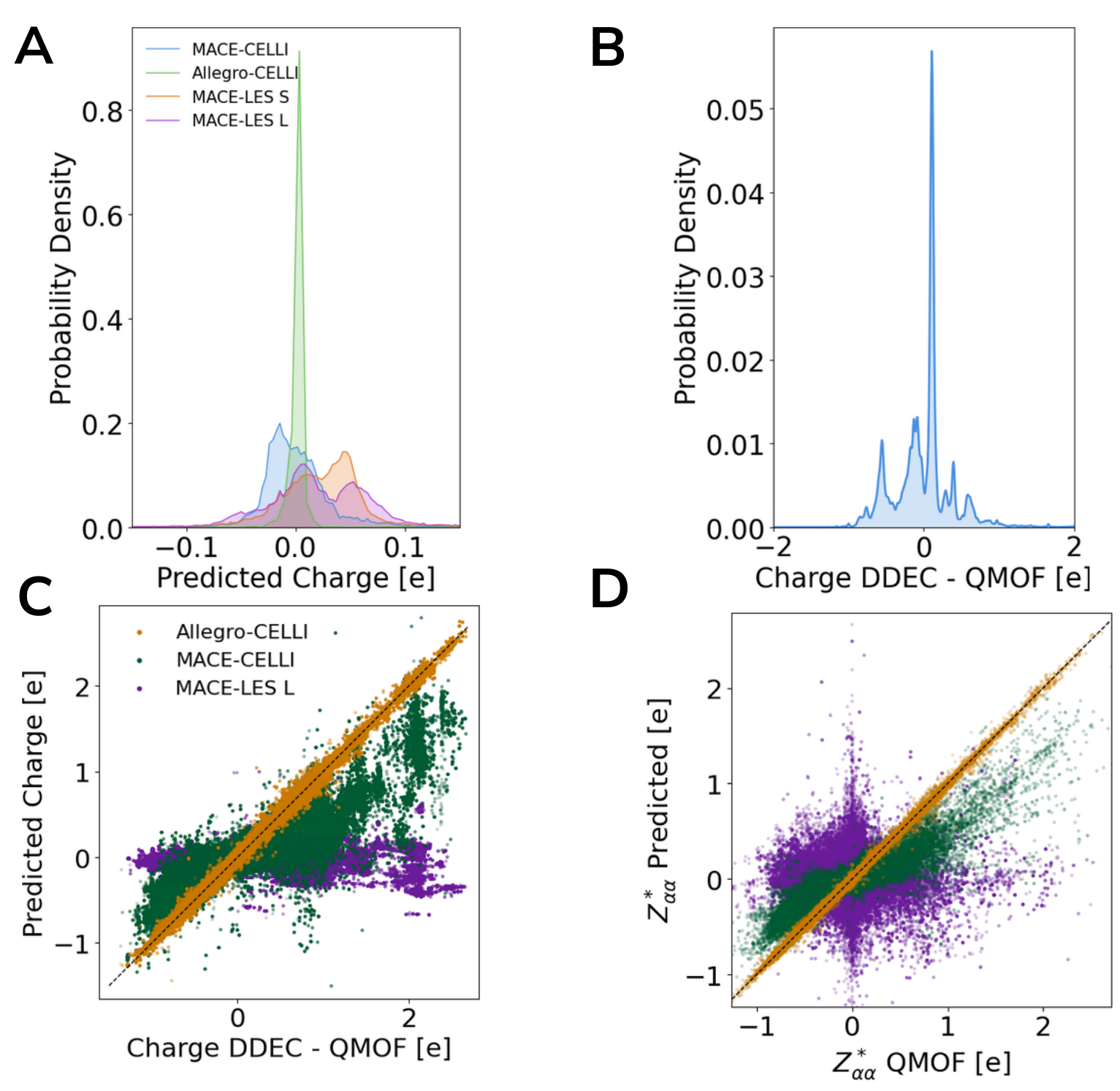}
    \caption{  \textbf{Inferring Charge distribution in MOFs.}
    Probability density of charges inferred by different models on the ODAC25 dataset (\textbf{A}) 
   and reference DDEC6 charges from the QMOF dataset (\textbf{B}).  
   MACE-LES L and MACE-LES S differ in hyperparameters, with MACE-LES L being more robust.
   Probability densities were obtained using Gaussian kernel density estimation in SciPy \cite{2020SciPy-NMeth}.   
    Parity plot of charges (\textbf{C}) and born effective charges (\textbf{D}) for models trained on a random split of the QMOF dataset (LES was excluded from this plot for clarity and moved to Supporting Information). As in ODAC25, LES was not able to recover the charge distribution. Despite MACE-CELLI being fitted to reference charges, a sufficiently low weight for the charge in the loss function allowed it to deviate from the reference charges. For details and additional plots, see Supporting Information section 9.}

    \label{fig:charge_odac}
\end{figure*}

In contrast to the other two datasets, ODAC25 does not provide reference partial charges, meaning that CELLI has no target charge values to fit. Although it is technically possible to use CELLI without reference charges, there is no guarantee that the inferred charges would accurately represent the underlying electrostatics of the system. Importantly, in this dataset, EFA and CELLI do not lead to improvements in either forces or energies (see Figure \ref{fig:ODAC}). A pronounced increase in error is observed for MACE-based models in the size split, where their performance is significantly poorer than in the other benchmarks. A closer inspection reveals that CELLI-based models were unable to infer meaningful charges in the absence of suitable reference data and effectively predicted zero charge for the majority of atoms (see Figure \ref{fig:charge_odac} \textbf{A}), even on a random split. Consequently, unlike in earlier benchmarks, Allegro with EFA outperforms CELLI across most benchmarking subsplits. Therefore, we recommend using CELLI only when reference charges are available, at least for complex systems such as MOFs.

The issue of relying on partial charges to model electrostatic interactions is well known \cite{Shaidu2024}, as charge-based schemes require predefined charges and their accuracy depends strongly on the chosen charge partitioning method, which often yields significantly different results \cite{Mei2015-mw}. To address this, several approaches have been proposed to bypass the need for reference charges by inferring them directly from forces and energies, such as the LES framework \cite{King2025-gt,Cheng2025-dw, kim2025universalaugmentationframeworklongrange}. \citeauthor{King2025-gt} demonstrated that LES can successfully infer charges for small systems (e.g., polar dipeptides from the SPICE dataset \cite{Eastman2023-xe,King2025-gt}); however, to our knowledge, it has never been tested on MOFs, which exhibit significantly more complex electrostatic environments than small biomolecules.

The MACE-LES model trained on the random split achieved MAE of 8.7 meV/atom and a force MAE of 24.6 meV/Å, meaning that the force error is around 50\% higher than that of the baseline MACE model, whereas energy error is 50\% lower (we also tested whether we can reproduce results reported before by \citeauthor{King2025-gt} to ensure that this is due to performance of the MACE framework rather than technical issues on our side for details see Supporting Information section 8). Further investigation shows that LES is also unable to infer correct charges without reference, and similarly to CELLI, it  predicts most partial charges to be almost zero. To further validate this result, we trained an additional MACE-LES model on the QMOF dataset, which includes DFT-derived partial charges and allows direct comparison to reference values. As shown in Figure \ref{fig:charge_odac}, the same behavior was observed: LES again failed to infer meaningful charges and collapsed to predicting near-zero values. Although LES frequently assigns charges opposite in sign to the reference values, this is not inherently problematic-Coulomb’s law is symmetric under global charge inversion, meaning forces and energies would remain unchanged if all charges were flipped. What is problematic is the lack of consistency, as LES sometimes predicts correct signs and sometimes their opposites. Based on this, we attribute better performance on energies of MACE-LES compared to baseline MACE are likely due to discrepancies in MACE implementations rather than properties of LES. Namely, baseline MACE was trained in JAX \cite{marioMACEjax} and MACE-LES in PyTorch \cite{ilyes_batatia_2025_16748079}. However, given that the ODAC25 subsets used in this study are relatively small, we cannot rule out the possibility that CELLI or LES may recover correct charge distributions when trained on sufficiently large datasets. Nevertheless, these results indicate that when no partial charges are available, other approaches are necessary, as none of the schemes achieved a significant improvement, except for the Allegro–EFA model. We would also like to highlight that in this work, accurate partial charge prediction is not treated as a goal in itself, but rather as a diagnostic indicating whether a model is able to infer meaningful long-range electrostatics, with the near-zero charges predicted by LES and CELLI in the absence of reference data signaling a failure to capture such effects in MOFs.

\section*{Conclusions}

Overall, our results establish a clear framework for evaluating long-range MLIPs on complex, chemically diverse systems and demonstrate that physically grounded treatments of electrostatics are essential for robust generalization. The presented biased split methods provide a robust test for generalizability to out-of-distribution samples, which is necessary for practical MLIP applications. They are also more broadly applicable than alternative functional group–based biased splits (for example, training on all structures except aliphatic hydrocarbons and testing on those), as they do not rely on predefined chemical heuristics. The proposed methods could capture differences arising from coordination environments, pore geometries, or topology features that are especially important in the context of MOFs, where structural diversity extends far beyond simple chemical fragments. Thus, they provide a more flexible and less assumption-driven assessment of MLIP generalizability and can be easily applied to other datasets with minimal modifications. 

Nevertheless, the reliance on SOAP descriptors may partly explain the limited decrease in performance observed for the ODAC25 subset, as this representation likely restricts the effectiveness of our splitting strategy. SOAP may fail to provide a sufficiently meaningful representation of certain atomic environments, resulting in a biased split that, in practice, is not very different from a random one. Additionally, SOAP descriptors are inherently short-ranged, and therefore may not fully distinguish structures that differ primarily in long-range electrostatics, potentially limiting how strictly these splits probe out-of-distribution behavior with respect to long-range physics. Nevertheless, if long-range interactions are not properly captured, models may compensate by overfitting short-range contributions, which can degrade short-range generalizability across chemically diverse environments. Future work could explore alternative representations, such as learned embeddings, electronic descriptors \cite{De_Armas-Morejon2023-oa}, or graph similarity measures \cite{Shiokawa2025-wq}, to construct more reliable biased splits. However, results from the QMOF dataset suggest that this limitation is not universally detrimental, and that SOAP-based splits can still expose meaningful performance differences when applied to sufficiently diverse datasets.

The performed benchmark tests clearly show that the incorporation of long-range schemes is necessary to achieve accurate and generalizable MLIPs. Especially in the QMOF benchmark, both Allegro and MACE benefited from the introduction of either EFA or CELLI, and only models based on CELLI exhibited good results in all three biased split benchmarks. We also show that the same cannot be achieved by increasing message passing layers. Furthermore, all variants share identical hyperparameters and nearly identical parameter counts, hence the observed improvements cannot be attributed to increased model capacity and instead arise from explicit long-range corrections.
Although total charge embeddings can recover much of CELLI’s performance on the OMOL25 dataset, they lack true long-range capabilities and generalize poorly in size-based extrapolation, indicating that embeddings alone cannot substitute for physically grounded treatments. In contrast, charge-based methods such as CELLI are inherently suited to modeling charged systems without relying on auxiliary embeddings, making them more robust and transferable, while embeddings should be viewed as complementary enhancements rather than replacements. 

Interestingly, our results reveal some discrepancies with earlier studies that introduced several of the long-range methods evaluated here. For instance, \citeauthor{fuchs2025learningnonlocalmolecularinteractions} reported that adding CELLI to MACE does not substantially improve performance over the baseline model, whereas in our out-of-distribution test cases, we observe clear gains. Similarly, our results indicate that LES is not able to infer charges from forces and energies alone, in contrast to conclusions drawn from smaller and less complex systems \cite{King2025-gt}. These differences can likely be attributed to two main factors: our use of out-of-distribution splits and the fact that we apply these models to more challenging systems than those commonly considered in similar studies. In addition, LES relies on a fixed smearing parameter \cite{grasselli2026longrangeelectrostaticsatomisticmachine}, which may further restrict transferability across chemically diverse environments \cite{zaverkin2025transferable}. This also highlights a broader issue in the development of long-range methods: they are often benchmarked on datasets containing relatively small molecules or systems in which long-range interactions are weak or negligible. Our results indicate that such evaluations can be misleading. Long-range schemes should instead be tested on more challenging systems, such as MOFs, where the electrostatic environment is highly complex and long-range contributions play an important role.

We should also note that long-range interactions can extend beyond pairwise Coulomb terms due to polarization and higher-order multipole effects \cite{ssp9-7s81}. Those effects are not explicitly modeled in the present CELLI and LES implementations. However, they can, in principle, capture polarization through charge prediction at each timestep. Future work could explore replacing QEq with polarizable charge equilibration schemes such as PQEq \cite{Gao2025,10.1063/1.4978891, grasselli2026longrangeelectrostaticsatomisticmachine}, or employing more expressive long-range kernels, including sum-of-Gaussians approaches (SOG-Net) \cite{ssp9-7s81}, to better capture these contributions. In addition, EFA may also capture higher-order and polarization effects. While detailed inference-time benchmarking is beyond the scope of this work, existing studies indicate that the considered long-range approaches do not significantly increase evaluation time \cite{fuchs2025learningnonlocalmolecularinteractions,frank2024euclideanfastattentionmachine}, further suggesting that they can be routinely integrated into MLIPs.

Our results also provide a cautionary perspective on the limits of inferring charges solely from energy and force. While CELLI delivers substantial improvements when reliable reference charges are available, its performance collapses in their absence, as it fails to recover meaningful electrostatic information and degrades overall accuracy. LES, which was designed with charge inference in mind, exhibits a similar breakdown in MOFs, as it consistently converges to near-zero or inconsistently signed charges across both ODAC25 and QMOF, suggesting that charge inference becomes unreliable once the electrostatic environment becomes too complex. Collectively, these findings suggest that charge-inference schemes, while appealing, are not yet robust enough for systems with complex long-range physics such as MOFs. Although none of the tested models were able to correctly infer charges without reference data, it is possible that more robust architectures or optimized hyperparameter choices could achieve this; however, a systematic investigation of such configurations is beyond the scope of this study. A potential solution to this issue may involve pretraining on reference charges followed by training solely on forces and energies, resulting in a scheme at least partially decoupled from the chosen charge partitioning method. Other promising directions could include modifying the loss function, introducing additional constraints into the learning process, predicting the whole electron density, or using methods that skip charges altogether, like EFA. Ultimately, the most appropriate long-range strategy depends on whether access to accurate partial charges is beneficial. CELLI is preferred when reliable electrostatics and charge‐transfer are important and difficult to infer directly from data, whereas charge‐free approaches such as EFA can be used for systems where explicit electrostatic information is not needed. 

\section*{Author contributions}
M.S. performed the experiments, analyzed the results, and wrote the manuscript. J.Z. supervised the project, reviewed the manuscript, provided resources, and acquired funding.

\section*{Conflicts of interest}
The authors declare no financial or non-financial conflicts of interest. The authors have contributed to the development of the CELLI method used in this study. 
\subsection*{Data Availability}

The datasets used in this study are publicly available to download. Our experiments utilized QMOF v1.4, OMOL25 (accessed June 20, 2025), and ODAC25 (accessed September 3, 2025). Details on the creation of utilized subsets of ODAC25 and OMOL25 datasets can be found in the Supplementary Information section 1.

\subsection*{Code Availability}
The software \texttt{chemtrain} used to train MLIPs is publicly available at \url{https://github.com/tummfm/chemtrain} version 0.1.0. The CACE and MACE models with LES were trained using code available at \url{https://github.com/ChengUCB/les/tree/main}, \url{https://github.com/BingqingCheng/cace}, and \url{https://github.com/ChengUCB/mace} ( accessed 09.2025). All code utilized in this work including requirements is available at \url{https://github.com/msan9908/MLIP_generalizability}.

\section*{Acknowledgements}

Funded by the European Union. Views and opinions expressed are, however,
those of the author(s) only and do not necessarily reflect those of the European Union or the European Research Council Executive Agency. Neither the European Union nor the granting authority can be held responsible for them. This work was funded by the ERC (StG SupraModel) - 101077842.



\balance


\bibliography{rsc} 
\bibliographystyle{rsc} 
\end{document}